\documentclass[12pt]{article}
\usepackage{epsfig}

\def\beq{\begin{equation}}
\def\eeq#1{\label{#1}\end{equation}}
\def\eeqn{\end{equation}}

\def\beqa{\begin{eqnarray}}
\def\eeqa#1{\label{#1}\end{eqnarray}}
\def\eeqan{\end{eqnarray}}

\let\bar=\overbar

\def\Dslash{\not{\hbox{\kern-4pt $D$}}}
\def\dslash{\not{\hbox{\kern-2pt $\del$}}}

\def\msb{{\bar{\ssstyle M \kern -1pt S}}}

\def\Title#1{\begin{center} {\Large {\bf #1} } \end{center}}

\begin{document}

\Title{Mini-Bias and Underlying Event Studies at CMS}

\bigskip\bigskip


\begin{raggedright}  

{\it Yuan CHAO\index{CHAO, Y.}\\
Department of Physics\\
National Taiwan University\\
10617 Taipei, TAIWAN}
\bigskip\bigskip
\end{raggedright}

\section{Introduction}

The Tevatron experiments provide us very good information for the quantum
chromodynamics (QCD) modelings of event generators. However, in the LHC era,
the collisions is in different region of phase-space from that of Tevatron.
A na\"ive rescaling of cross-sections will not work. The current modeling of
non-trivial interplay of perturbative and non-perturbative aspects based on
Tevatron data have large discrepancies when extrapolated to the LHC energy.
This study is devoted to the proton-proton dynamics exploration:
discriminating among different QCD Monte Carlo models.

\section{The Minimum Bias events and triggers}

From the experimental point of view, a minimum bias events (MB) correspond
to a non-single diffractive inelastic interaction. Meanwhile a totally
inclusive trigger, or called zero bias trigger, corresponds to a randomly
reading out from the detector whenever a collision is possible. The later
is only efficient when the luminosity is high enough such that a reasonable
probability of collisions occur during a bunch crossing.

For events with  one and only one collision occurs is called ``ideal data'',
which have no piled-ups, during a bunch crossing. This kind of events is
important for underlying events (UE) study as no influence from multiple
p-p interactions.

The average number of collisions, constituting the pile-up, per beam bunch
can be described as:
\[
<N_{int.}> = L_{inst.} \times \sigma / f_{rev.}
\]

The cross-section $\sigma$, includes both elastic and inelastic,
is $\sim 100$ mb at LHC with mostly low $p_T$ particles and low multiplicity.
So the averaged number is about 35 events per bunch-crossing for luminosity
L=1034 $/cm^2/sec$. 

\subsection{Minimum-Bias Trigger}
The MB trigger in CMS is using the Hadron Forward (HF) calorimeter.
It has a geometrical coverage from 3 to 5 in absolute value of pseudo-rapidity
$\eta$ and consists of 18 wedges per side with tower size $0.175\times 0.175$ in
$\eta$ and $\phi$. A trigger based on minimum 10 towers with energy threshold
grater than 1.4 GeV gives $90\%$ efficiency.

\subsection{Charged Hadron Spectra}
A study on the charged hadron spectra on MB events has been performed.
The measured differential yields of unidentified charged particles as
well as pions, kaons and protons are shown as a function of $p_T$ and in
narrow $\eta$ bins. Tsallis function fits are also superimposed. The results
are using a sum of both positive and negative charged particles and assuming
symmetric $\eta$ bins.
\begin{center}
\epsfig{file=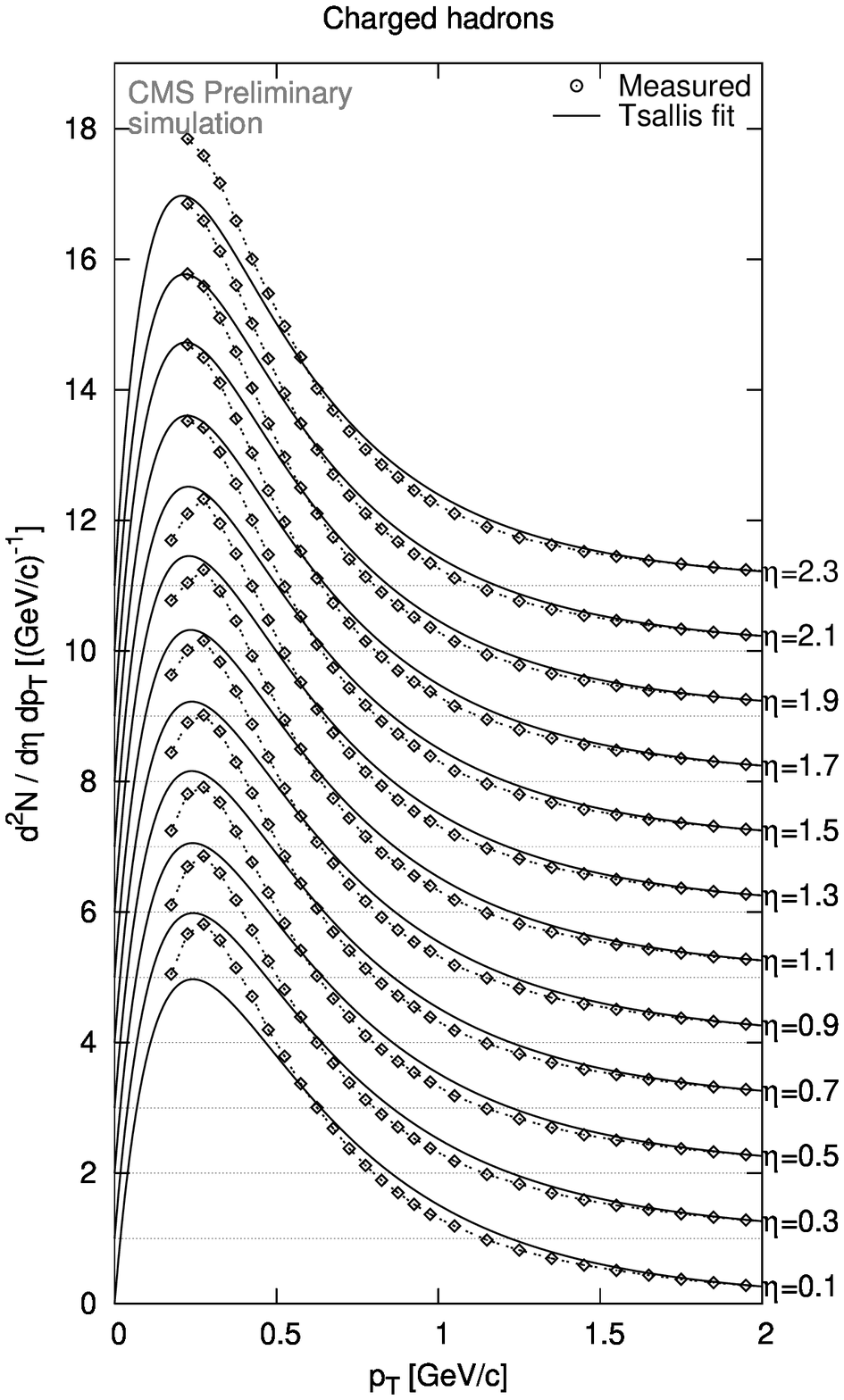,height=1.7in}
\epsfig{file=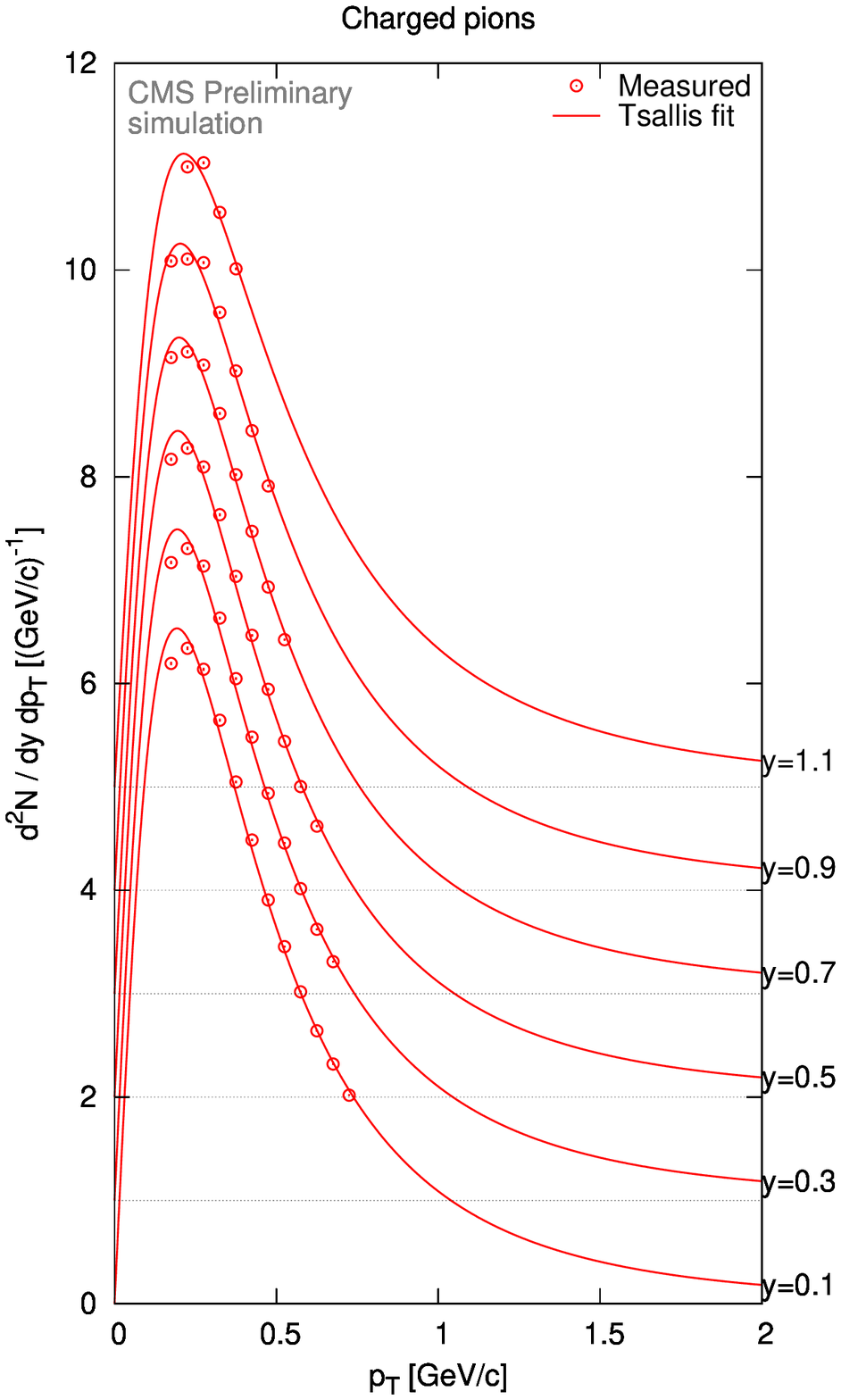,height=1.7in}
\epsfig{file=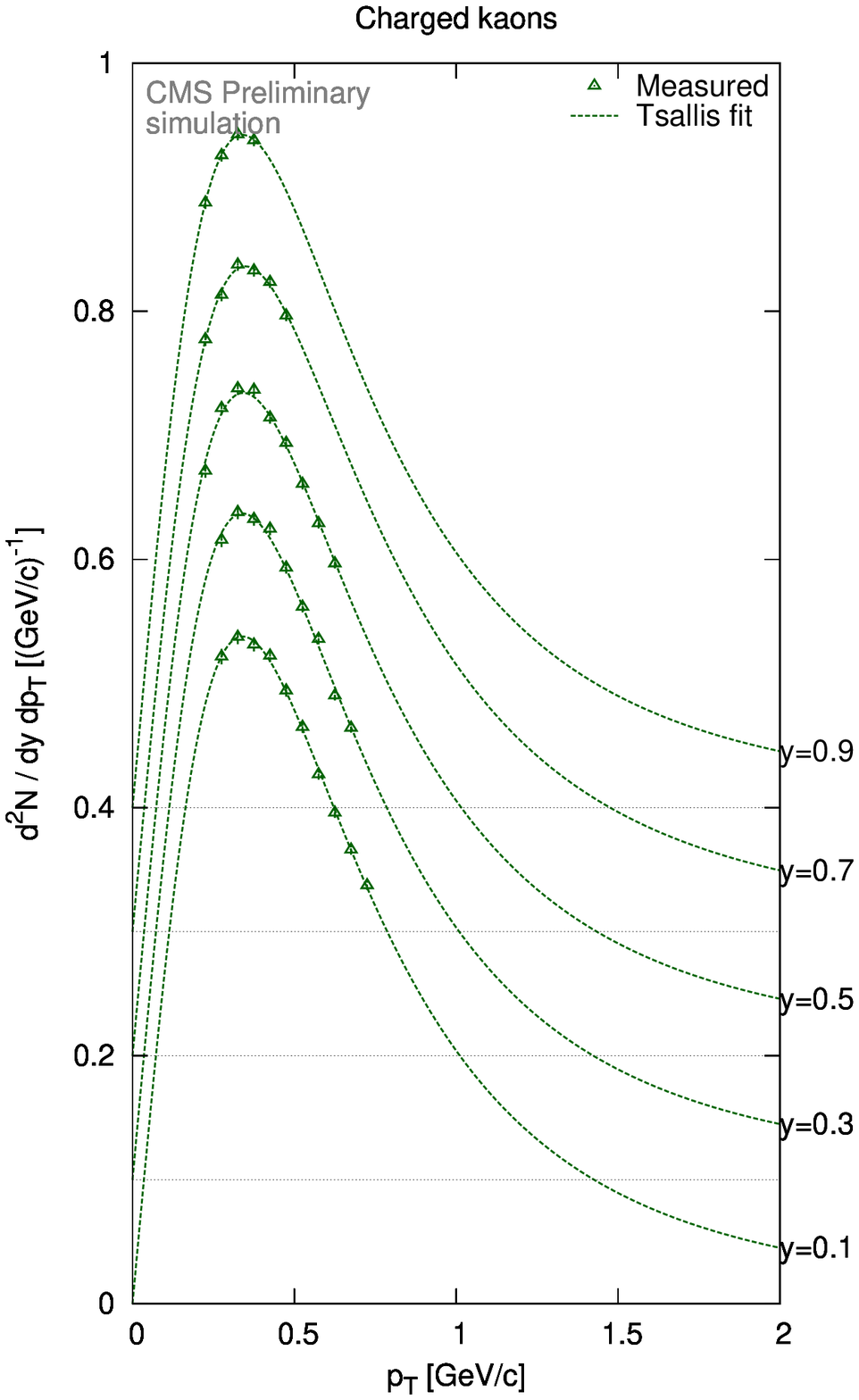,height=1.7in}
\epsfig{file=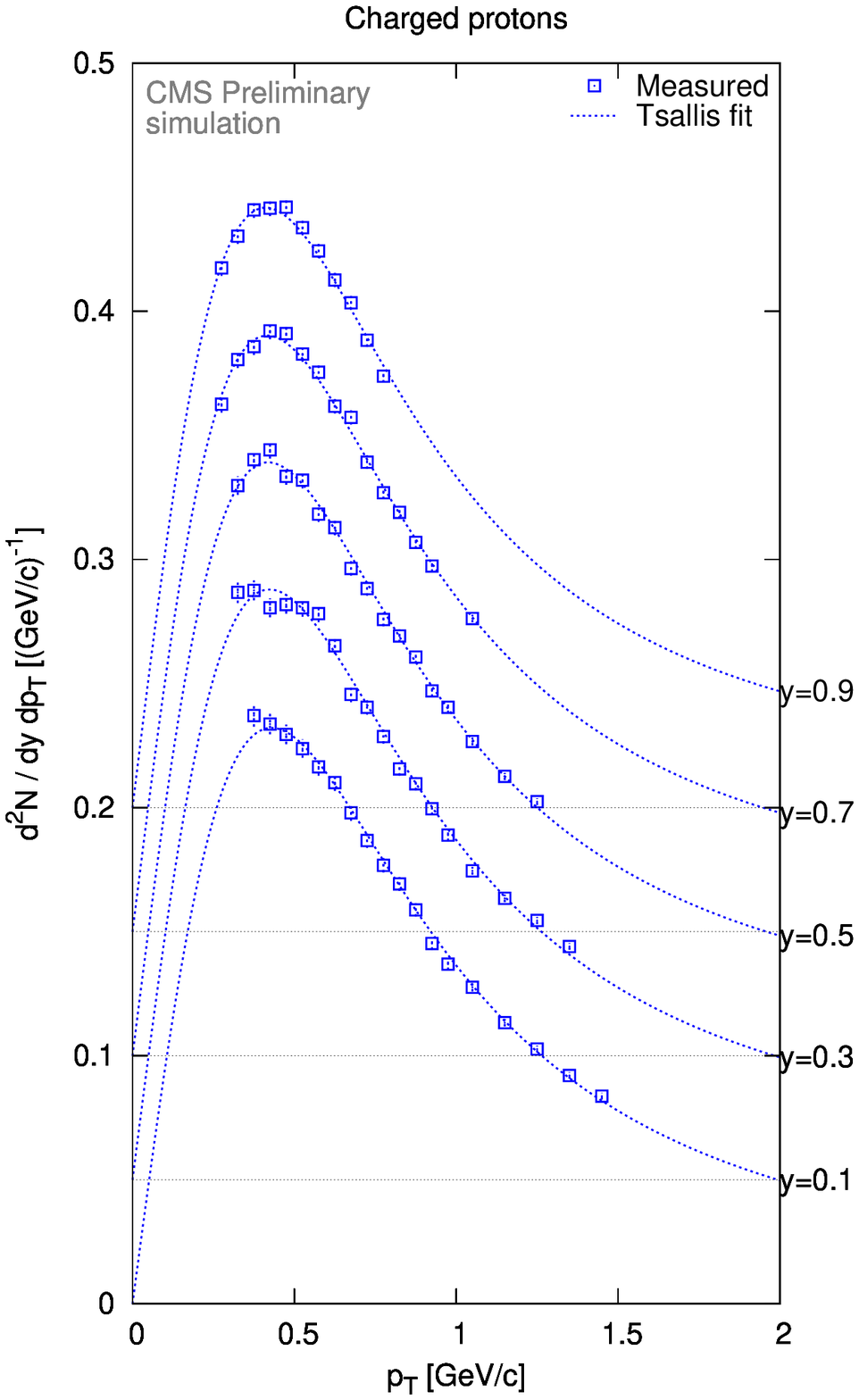,height=1.7in}
\end{center}

\subsection{Energy dependency}
The density of charged hadrons at $\eta \sim 0$ follows the trend from lower
energies, which is linear in $\log\sqrt{s}$. CMS expected to see an averaged
4.2 charged hadrons in the central region. Meanwhile the averaged transverse
momentum of charged hadrons at lower energies is described by a quadratic
function in $\log\sqrt{s}$. The expected average $p_T$ is about 0.7 GeV/c
for CMS.
\begin{center}
\epsfig{file=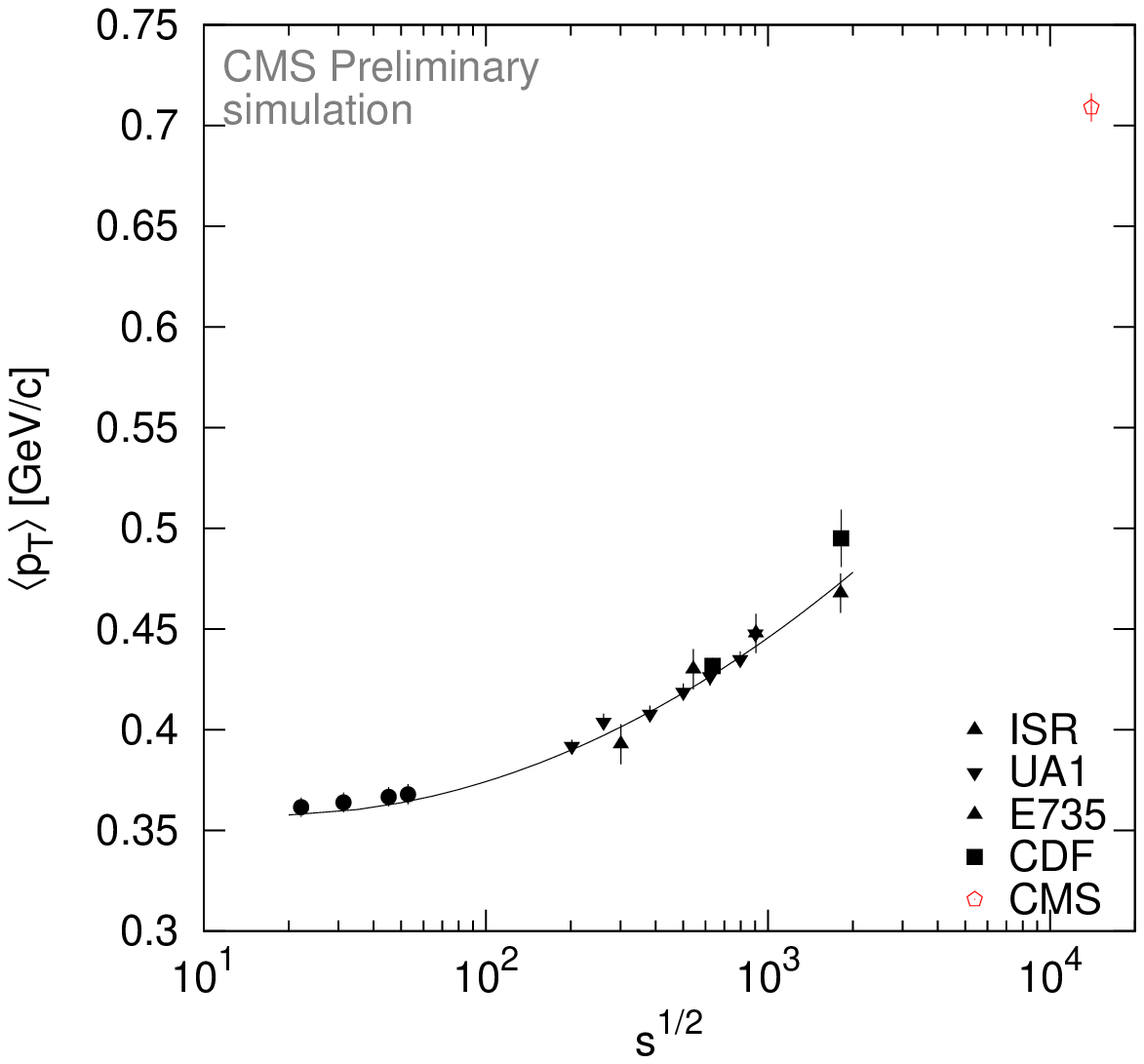,height=1.2in}
\epsfig{file=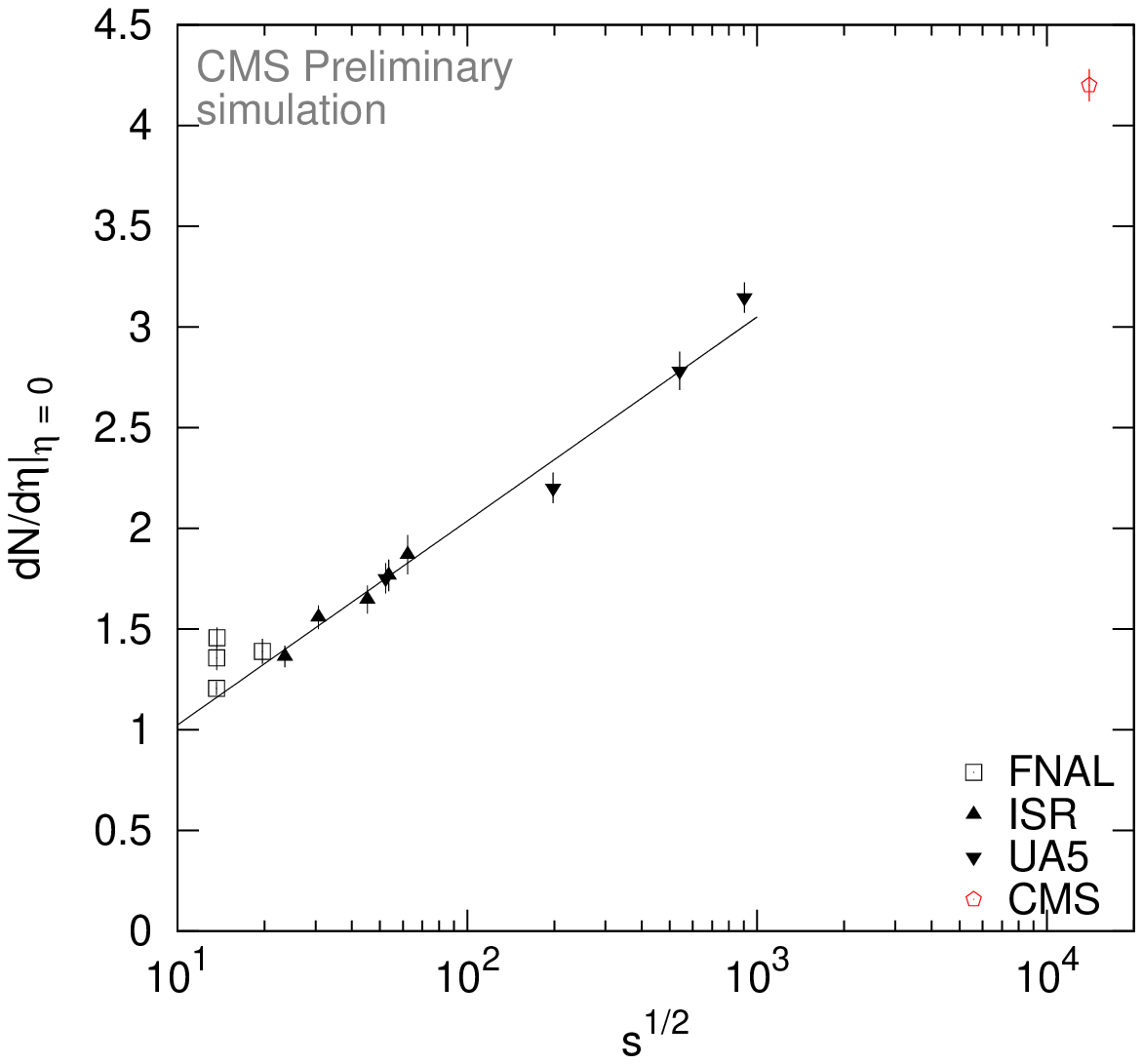,height=1.2in}
\end{center}

\section{Underlying Events}

The ``underlying events'' (UE) is everything in a single proton-proton
interaction except for the hard scattering component. It's not a minimum-bias
event on top of the hard process. What happens to the beam remnants after the
hard scattering is an important issue. The UE has the same production vertex
so it's tied to the process of interest. Its activity also grows with the
process energy scale as a ``pedestal effect''. 

UE phenomenology has been studied with CDF data using ``charged jet'' from
iterative cone algorithm on mass-less tracks. With various energy scale in $p_T$
of the charged jet, multiplicity and $p_T$ density has been studied in the
transverse region. As we can easily tell from the following plots, the
activities in this region have less influence from the hadron event.

Clear dependency on $p_T$ can also be seen for charged jets in the toward
region.

\subsection{Underlying Event Models}
The UE event modeling has both non-perturbative and perturbative aspects.
The former ones include initial-state radiation (ISR), final-state radiation
(FSR) and beam remnants together with multiple parton-parton interaction,
while the later part have the following considerations: allowing more than
one parton-parton interaction per pp-scattering (MPI); regularizing QCD
two-to-two cross section cut-off on $p_T$, e.g. Pythia DW and DWT tunes;
variable impact parameter models; and color reconnection models, e.g.
Pythia S0. However, the current models based on Tevatron data have different
behavior when extrapolating to LHC.

From an experimental point of view, one can use a topological structure
of hadron-hadron collisions to probe the UE activities out of the hard
scattering component. The activities near the transverse plane to the jet
direction has the smallest influence from the hard scattering and provides
the most sensitivity to the UE contributions.

\subsection{Results}
Herwig (without MPI), Pythia tunes DW, DWT, and with MPI), all predict
observed results from the Tevatron. One can discriminate between these
scenarios at LHC energies by looking at the density of charged particles
$d N/d \eta d\phi$ and the momentum density $d p_T^{sum}/
d \eta d\phi$ in the transverse region.

Density of charged particles and momentum of the leading charged particle
jet using a track reconstruction threshold of 0.9 GeV/c is shown. Data points
from different triggers are superimposed (Minimum Bias, JET20, JET60, JET120)
and correspond to the corrected reconstruction level profiles using tune data).
The lines correspond to the different generator level tunes: DW, DWT, S0 and
HERWIG.

By lowering the $p_T$ threshold to 0.5 GeV/c it is possible (largely due to MB
data) to distinguish between DW/DWT and S0 as well.
\begin{center}
\epsfig{file=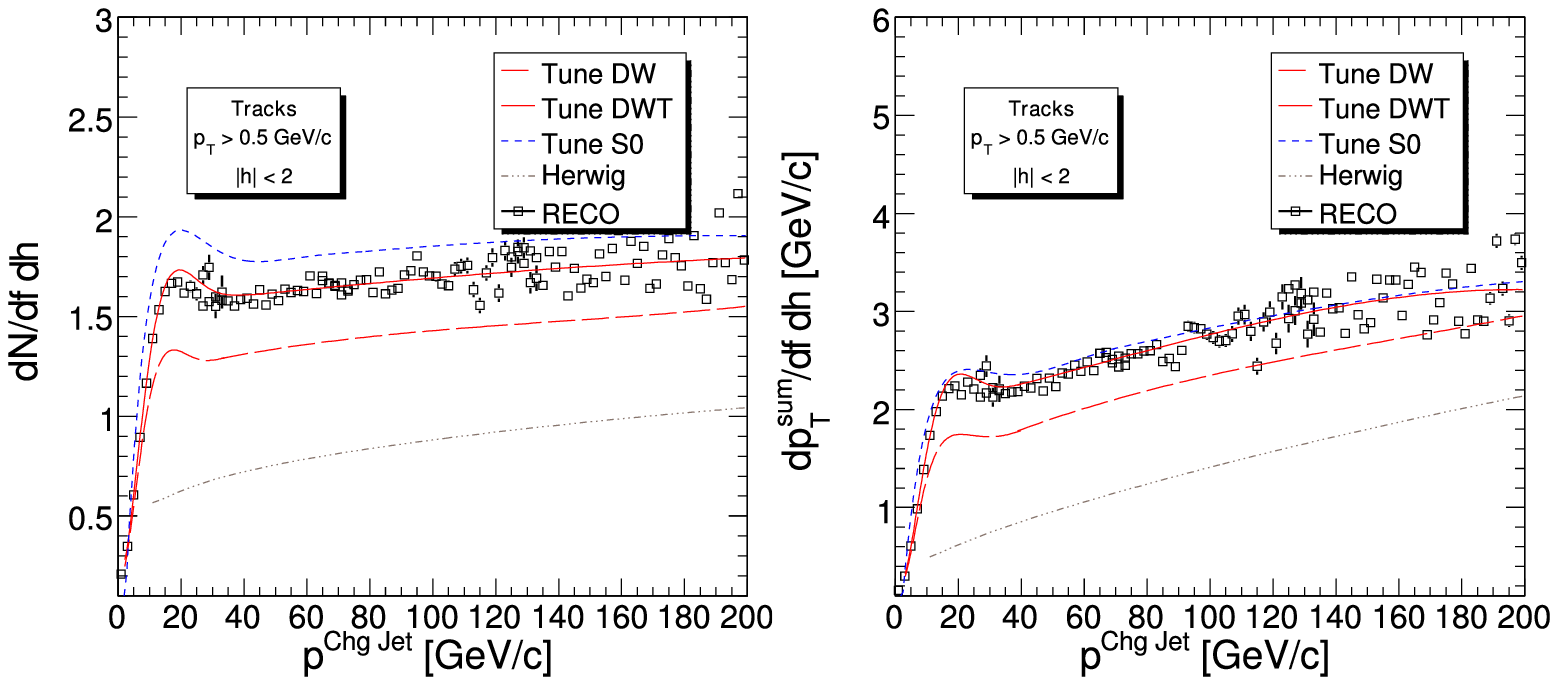,height=1.2in}
\epsfig{file=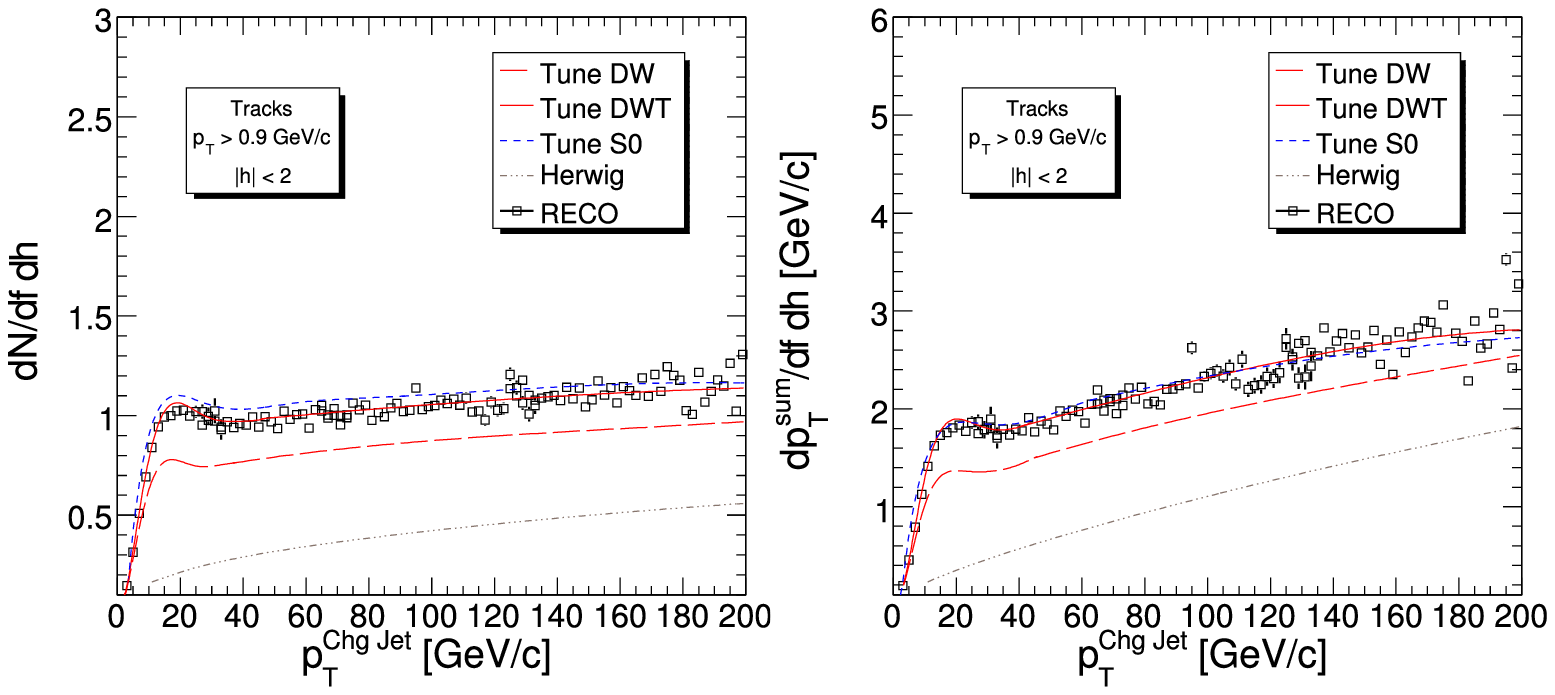,height=1.2in}
\end{center}

With 100 pb$^{-1}$ of data it should be possible to discern between Herwig DW,
and the two other Pythia tunes (DWT, SO) using a $p_T$ threshold of 0.9 GeV/c.
\begin{center}
\epsfig{file=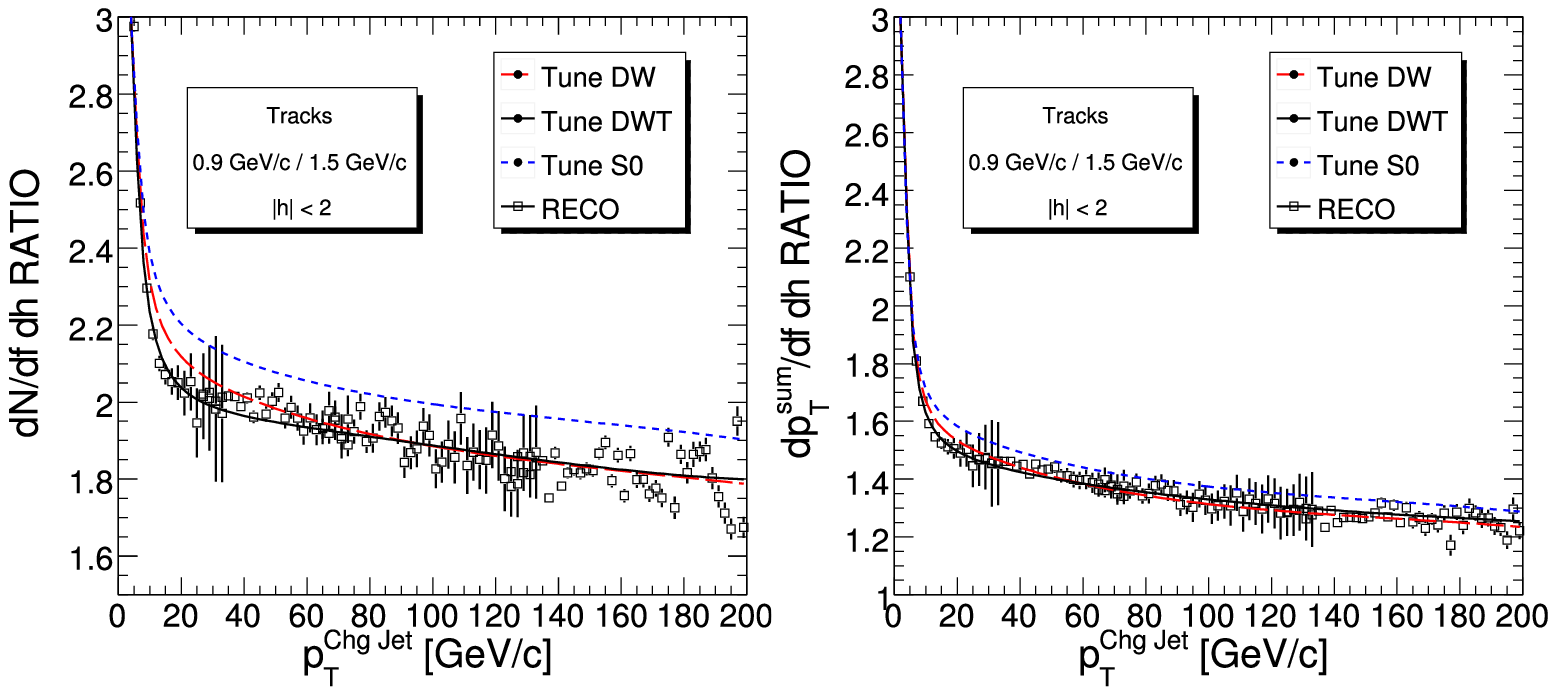,height=1.2in}
\end{center}


\section{Conclusion}

The UE study can help us to discriminate between various QCD models which
will facilitate the improvement and tuning of Monte Carlo models at LHC
start-up. It will also open prospects for the exploration of QCD dynamics
in proton-proton collisions at 10 TeV.

\def\Discussion{
\setlength{\parskip}{0.3cm}\setlength{\parindent}{0.0cm}
     \bigskip\bigskip      {\Large {\bf Discussion}} \bigskip}
\def\speaker#1{{\bf #1:}\ }
\def\endDiscussion{}



 

\begin{thebibliography}{99}

\bibitem{LHC}
{\it ``LHC Machine''},
Lyndon Evans and Philip Bryant (editors),
JINST 3 S08001 (2008)

\bibitem{CMS}
{\it ``The CMS experiment at the CERN LHC''}
The CMS Collaboration, S Chatrchyan {\it et al.}, JINST 3 S08004 (2008)


\bibitem{ch_had}
{\it ``Measurement of charged hadron spectra in proton-proton collisions at
$sqrt{s}$ = 14 TeV''}, PAS-QCD-07-001

\bibitem{MBUE}
{\it ``Measurement of the Underlying Event in Jet Topologies using Charged
Particle and Momentum Densities''}, PAS-QCD-07-003


\end{thebibliography}
\end{document}